\documentclass[12pt]{article}

\usepackage[utf8]{inputenc}
\usepackage[T1]{fontenc}
\usepackage[margin=1in]{geometry}
\usepackage{graphicx}
\usepackage{amsmath}
\usepackage{amssymb}
\usepackage{authblk}
\usepackage{caption}
\usepackage{subcaption}
\usepackage{booktabs}
\usepackage{setspace}
\usepackage[numbers,sort&compress]{natbib}
\usepackage[colorlinks=true,linkcolor=blue,citecolor=blue,urlcolor=blue]{hyperref}
\usepackage{xcolor}
\usepackage{lineno}

\graphicspath{{./}}

\title{\textbf{Towards Precision Cardiovascular Analysis in Zebrafish:\\
The ZACAF Paradigm}}

\author[1]{Amir Mohammad Naderi}
\author[2]{Jennifer G. Casey}
\author[1]{Mao-Hsiang Huang}
\author[3]{Rachelle Victorio}
\author[3]{David Y. Chiang}
\author[3]{Calum MacRae}
\author[1,4,5,$*$]{Hung Cao}
\author[2,$*$]{Vandana A. Gupta}

\affil[1]{Department of Electrical Engineering and Computer Science,
          University of California, Irvine, CA}
\affil[2]{Division of Genetics, Department of Medicine,
          Brigham and Women's Hospital, Harvard Medical School,
          Boston, MA, USA}
\affil[3]{Cardiovascular Medicine, Department of Medicine,
          Brigham and Women's Hospital, Harvard Medical School,
          Boston, MA, USA}
\affil[4]{Department of Biomedical Engineering,
          University of California, Irvine, CA}
\affil[5]{Sensoriis, Inc., Edmonds, WA}
\affil[$*$]{Co-corresponding authors}

\date{}

\begin{document}
\maketitle

\begin{abstract}
Quantifying cardiovascular parameters like Ejection Fraction (EF) in zebrafish
for a host of biological investigations has been extensively studied.
Since current manual monitoring techniques are time-consuming and fallible,
several image processing frameworks have been proposed to automate the process.
Most of these works rely on supervised deep-learning architectures.
However, supervised methods tend to be overfitted on their training dataset.
This means that applying the same framework to new data with different imaging
setups and mutant types can severely decrease performance.
We have developed a Zebrafish Automatic Cardiovascular Assessment Framework
(ZACAF) to quantify the cardiac function in zebrafish.
In this work, we further applied data augmentation, Transfer Learning (TL),
and Test Time Augmentation (TTA) to ZACAF to improve the performance for the
quantification of cardiovascular function in zebrafish.
This strategy can be integrated with the available frameworks to aid other researchers.
We demonstrate that using TL, even with a constrained dataset, the model can be
refined to accommodate a novel microscope setup, encompassing diverse mutant types
and accommodating various video recording protocols.
Additionally, as users engage in successive rounds of TL, the model is anticipated
to undergo substantial enhancements in both generalizability and accuracy.
Finally, we applied this approach to assess the cardiovascular function in
\textit{nrap} mutant zebrafish, a model of cardiomyopathy.
\end{abstract}


\section{Introduction}

Zebrafish have emerged as the vertebrate model to study development and diseases
in the past few decades due to embryonic transparency, high genomic conservation
between zebrafish and humans, and ease of genetic manipulations.
Identification and function of novel or disease-associated genes can be investigated
by several forward or reverse genetic approaches.
Following the proposal of ZACAF~\cite{naderi2021}, numerous research groups
expressed interest in its utilization.
We also extended the ZACAF concept to zebrafish echocardiography videos~\cite{huang2023}.
However, it became evident that the deep learning model at the core of the framework,
originally trained on a limited dataset with a distinct imaging setup, including recordings
from the TTNtv mutant type~\cite{wheeler2005}, posed challenges when applied to new data
received from various groups.
This underscored the necessity for an enhanced framework capable of accommodating
diverse transgenic lines and providing a robust tool for the precise study of zebrafish
cardiovascular parameters in genetic investigations.
To fill this gap, we present an improved platform that refines ZACAF and assesses
its adaptability with a new transgenic line of zebrafish, showcasing its efficacy in a
broader context.
This upgraded framework offers user-friendly features, addressing the limitations of
the initial model and establishing itself as a reliable tool for the exploration of
zebrafish cardiovascular parameters, especially within the context of different
transgenic lines pertinent to genetic studies.

Nebulin Related Anchoring Protein (NRAP) is a member of the Nebulin family of
proteins that functions as a thin filament chaperone to promote myofibril assembly
during early skeletal muscle development and is absent in the myofibrils in the
post-natal skeletal muscle~\cite{lu2008}.
However, aberrantly high levels of NRAP are present in skeletal muscle in human
myopathies and contribute to disease pathology by abnormal sequestration of
sarcomeric proteins~\cite{jirka2019a}.
Downregulation of NRAP rescues skeletal muscle structure and function in zebrafish
models of human myopathies~\cite{jirka2019a}.
Similarly, NRAP is required in cardiac muscle for myofibril assembly during heart
development~\cite{lu2011}.
However, transgenic overexpression of Nrap in the heart resulted in dilated
cardiomyopathy in mice~\cite{lu2011}.
Thus, the downregulation of NRAP provides therapeutic avenues in skeletal and
cardiac muscle disorders.
In recent years, rare loss of functional variants in NRAP have also been identified
in cardiomyopathy, but a lack of functional validation of these variants prevents a
strong casual association of NRAP with the disease state.
Moreover, the presence of a homozygous truncating variant (rs201084642) in both
affected and unaffected siblings suggests a low penetrance and allele risk~\cite{truszkowska2017}.
Biallelic NRAP variants were also identified in 11 unrelated probands with dilated
cardiomyopathy; however, a lack of functional analysis of these variants prevents
a strong genotype-phenotype correlation~\cite{koskenvuo2021}.
A novel loss of function mutation (c.259delC) in NRAP is also linked to left
ventricular noncompaction cardiomyopathy.
Functional modeling of this variant in zebrafish showed pericardial edema in a
fraction of mutants as well as wild-type zebrafish, suggesting the non-pathogenicity
of this variant~\cite{zhang2023nrap}.
While these studies suggest a potential link between NRAP deficiency and cardiac
defects, strong functional evidence is still lacking in classifying NRAP as a
cardiac disease gene.

Due to the potential therapeutic avenues of NRAP reduction in skeletal and cardiac
muscle myopathy, we aimed to investigate the cardiac effects, specifically EF and
Fractional Shortening (FS) in \textit{nrap} mutant zebrafish with ZACAF deep learning
model.
No significant differences in ventricle shape, EF, and FS were observed between
mutant and control genotypes in embryonic zebrafish.
Utilizing ZACAF for cardiovascular parameters, such as EF and FS from microscopic
videos, allows for a fully automated measurement of the mentioned indices~\cite{naderi2021}.
Furthermore, this trained model will provide a robust tool that can perform the
measurements on video recordings in other zebrafish models in a short time.
This approach will make tedious, inconsistent, and often inaccurate manual measurement
unnecessary.
This work demonstrates ZACAF's effectiveness in implementing data from different
research teams with different recording setups and phenotypes.
Additionally, considerations for recording and preprocessing the videos, orientation,
handling of the zebrafish, and tuning the architecture of the deep learning model
are discussed.
Furthermore, the benefits of data augmentation, TL, and TTA are investigated and
discussed in this work to improve robustness and applicability to new data from new
transgenic lines or research groups.

\subsection{Related Works}

The heart of a framework like ZACAF is its deep learning-based segmentation model
that inputs the video frames and outputs corresponding binary masks of the ventricle.
Among the most commonly utilized segmentation architectures are U-Net, FCN
(Fully Convolutional Network), and SegNet.
U-Net, known for its symmetric encoder-decoder architecture with skip connections,
has gained widespread adoption in biomedical image segmentation tasks due to its
ability to capture fine-grained spatial details effectively~\cite{ronneberger2015}.
FCN replaces fully connected layers with convolutional layers, enabling end-to-end
pixel-wise predictions and providing flexibility in handling images of variable
sizes~\cite{long2015}.
SegNet, similar to U-Net in its symmetric structure, utilizes an encoder-decoder
architecture without skip connections, relying on max-pooling indices for
up-sampling~\cite{badrinarayanan2017}.
Each of these architectures offers distinct advantages in different segmentation tasks,
catering to specific requirements such as spatial detail capture, flexibility, or
precise localization.

To date, the majority of existing studies have primarily focused on basic heart rate
detection methods, such as edge tracing~\cite{wessells2004}.
Nasrat et al.\ introduced a semi-automatic approach for quantifying fractional shortening
(FS) in zebrafish embryo heart video recordings~\cite{nasrat2016}.
Their software offers automated visual insights into end-systolic (ES) and end-diastolic
(ED) stages by displaying color-coded lines on a motion-mode display.
However, the manual marking of ventricle diameters during ES and ED stages, followed
by FS calculation, proves to be highly laborious, time-intensive, and prone to
inconsistencies when dealing with a large number of frames.
Akerberg et al.\ proposed a SegNet-based framework to automatically segment chambers
from videos and calculate ejection fraction (EF)~\cite{akerberg2019}.
Nonetheless, their approach relies on specific transgenic animals expressing
myocardial-specific fluorescent reporters and high-end fluorescence microscopes,
limiting its widespread applicability within the research community, especially for
those lacking access to such resources.
Furthermore, Huang et al.\ highlighted potential issues with transgenic fluorescence
protein expression leading to dilated cardiomyopathy~\cite{huang2000}, underscoring
concerns regarding the use of foreign proteins that may impact myocardial function.
Additionally, Akerberg et al.\ utilized frames from only four videos, raising concerns
about potential overfitting when video features such as fish position, lighting conditions,
or lens focus on the ventricle differ from the training set.
Zhang et al.\ proposed a U-Net based framework similar to ZACAF; however, similar
to Akerberg's study, they used a transgenic zebrafish line expressing reporters and
high-end fluorescence microscopes~\cite{zhang2021}.
Suryanto et al.\ used DeepLabCut for labeling the ventricle to facilitate automatic
zebrafish cardiac assessment~\cite{suryanto2022}.
DeepLabCut is a software toolbox that employs deep learning techniques to enable
markerless pose estimation and tracking in videos of animals or humans.
Nonetheless, this framework only marks 8 points on the ventricle, which limits the
resolution and accuracy of the predicted ventricle.
This will significantly limit the robustness of the framework, especially with mutant types.

\section{Methods}

\subsection{Experimental Animals}

Fish were bred and maintained using standard methods as described~\cite{westerfield1995}.
All procedures were approved by the Brigham and Women's Hospital Animal Care and Use Committee.
\textit{nrap\textsuperscript{sa42059}} zebrafish line was obtained from Zebrafish International
Resource Center (ZIRC) and genotyping was performed as previously described~\cite{casey2023}.
Zebrafish embryonic (0--2 days post fertilization) and larval stages (3--5 dpf) have been
defined as described previously~\cite{kimmel1995}.
\textit{nrap\textsuperscript{sa42059}} zebrafish harbor a point mutation resulting in a premature
stop codon (c.2106T$>$G); p.701*).
To evaluate the effect of Nrap deficiency on cardiac function, $+/+$, $+/-$ and $-/-$
siblings were analyzed at 2 dpf.

\subsection{Considerations for Image Acquisition}

\begin{figure}[htbp]
    \centering
    \includegraphics[width=\linewidth]{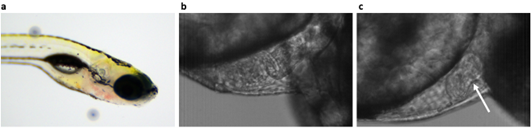}
    \caption{Comparing the visibility of the zebrafish ventricle using 5$\times$ (a) and
    10$\times$ zoom (b) and a pear-shaped ventricle with the arrow signifying the pear shape (c).
    As can be seen, in the 5$\times$ zoom image, the borders of the ventricle cannot be identified.
    However, in the 10$\times$ image, both chambers can be seen in the middle.
    The rightmost image is an example of a pear-shaped ventricle.}
    \label{fig:fig1}
\end{figure}

Zebrafish heart was visualized and imaged with a FastCam-PCI high-speed digital camera
(Photron, USA) with a frame rate of 250 frames per second (fps) attached to a Zeiss upright
microscope at 10X and processed with the FastCam-PCI image capture board (Figure~\ref{fig:fig1}).
To maximize video quality with reduced file size, videos were acquired in greyscale with a
resolution of $512 \times 480$ pixels for 4.35 seconds, which roughly captured 8 cardiac
cycles and 1,088 frames with a 0.004 shutter speed.
The placement of the fish under the microscope is another critical factor for imaging.
Firstly, in most literature for quantification of cardiac function using 2D imaging,
the ventricle is assumed to be an ellipsoid.
Hence, during image acquisition, it is important to make sure the fish is properly positioned
on its side.
Improper placement of the fish under the microscope can result in the ventricle having a
pear-shaped structure that results in inaccuracy in the quantification of EF.
Furthermore, it is important to know that the placement of the fish under the microscope
is a feature in deep learning models.
Uniform placement of the fish across the videos recorded for the dataset can affect the
model into only responding accurately to test videos with similar placement.
In order to train a robust framework less prone to the placement of the fish, data
augmentation must be used in the process of training the deep learning model.

\subsection{Cardiac Function Assessment}

Fractional shortening, one of the most crucial indices of ventricular contractility,
can be calculated using the ventricular diameters at end-diastolic (ED) and end-systolic
(ES) ($D_d$ and $D_s$) in the manner shown below:

\begin{equation}
    \text{FS} = \frac{D_d - D_s}{D_d}
    \label{eq:fs}
\end{equation}

Ventricular volumes need to be calculated to calculate stroke volume, EF, and cardiac
output (CO).
First, 2D still images are used to determine the long- and short-axis diameters ($D_L$
and $D_S$).
If the ventricle is prolate spheroidal in shape, the following volume formula can be
used~\cite{yalcin2017}:

\begin{equation}
    \text{Volume} = \frac{\pi}{6} \times D_L \times D_S^{2}
    \label{eq:vol_prolate}
\end{equation}

However, if we consider that the shape of the ventricle is unknown while having the 2D
shape of the ventricle, the volume can be calculated using the formula below:

\begin{equation}
    \text{Volume} = \frac{8}{3\pi D_L} \times A^{2}
    \label{eq:vol_area}
\end{equation}

In this formula, $A$ is the area of the segmented 2D ventricle~\cite{wisneski1981}.
It is noteworthy that these formulas assume ellipsoid shape for all ventricles, which
would increase the error in recordings with pear-shaped or abnormal ventricles.

Using the ventricle volumes at the ED (EDV) and ES (ESV), stroke volume (SV), or the
amount of blood pumped from the ventricle for each beat, can be calculated:

\begin{equation}
    \text{SV} = \text{EDV} - \text{ESV}
    \label{eq:sv}
\end{equation}

Ejection fraction is the amount of blood that exits the ventricle with each heartbeat
and may be calculated using the following formula:

\begin{equation}
    \text{EF}(\%) = \frac{\text{EDV} - \text{ESV}}{\text{EDV}} \times 100
                  = \frac{\text{SV}}{\text{EDV}} \times 100
    \label{eq:ef}
\end{equation}

The following formula can be used to compute CO from SV and heart rate
(HR)~\cite{yalcin2017}:

\begin{equation}
    \text{CO}\!\left(\frac{\text{nanoliter}}{\text{min}}\right) = \text{SV} \times \text{HR}
    \label{eq:co}
\end{equation}

HR is calculated using the interval between two identical consecutive points (i.e., ED
or ES) in the captured images.
For healthy zebrafish, the EF ranges between 50\% and 70\%, and it is one of the key
indicators in the identification of heart failure.
EF is a measure of how well the systolic ventricular pump works.
Higher SV and EF suggest stronger myocardial contractility.
As a result, the left ventricular EF is much lower in patients with heart failure.

\subsection{Automated Quantification of Cardiovascular Parameters Using ZACAF}

ZACAF is a U-Net-based deep learning architecture that was proposed to automate the
measurement of the above-mentioned indices~\cite{naderi2021}.
The proposed framework applies semantic segmentation to a sequence of frames from an
input video.
It then outputs a sequence of masks representing the ventricle in those frames.
The contour tool from OpenCV (an open-source computer vision library) automatically
measures the ventricle diameters for each extracted frame.
The ES and ED stages are depicted by the maximum and minimum measured areas of the
ventricle in various frames.
We can determine the SV, EF, and FS by measuring the ES and ED frames.
The interval between two ES (or ED) frames can also be utilized to calculate heart rate (HR).
It is assumed that the predicted ventricle will be elliptical.
By measuring the axis in the anticipated shape, it is possible to calculate the EF using
the ventricle volume (Eq.~\ref{eq:vol_prolate} above).
In this paper, the ZACAF model is applied and modified so it is more generalized for new
data, providing a more robust and accessible framework for researchers using zebrafish.

\subsubsection{Dataset}

In this study, a training dataset was constructed using raw microscopic videos of
zebrafish, comprising a total of 410 pixel-wise annotated images.
To create this dataset, 41 videos of the lateral view from 41 different 2-dpf zebrafish
were analyzed.
Specifically, 9 of these videos were obtained from the \textit{nrap} mutants, 19 from
the heterozygous zebrafish, and the remaining 9 from the wild-type siblings.
A visual validation was performed to select the videos to make sure the desired region
of interest was captured with adequate clarity on the ventricle.
The genotypes of these zebrafish were blinded until the final analysis to prevent bias
in the segmentation.
From each individual video, 10 sequential frames were extracted, resulting in a total
of 410 frames for the training set.
To select frames from each video, we opted to extract 10 frames starting from the
beginning, with intervals of 10 frames.
This approach ensures that the fish's heart, across various sizes within the heart rate
cycle, is adequately represented in the dataset.
Additionally, it ensures that the selected frames are suitably diverse.
Each training set consisted of an original frame extracted from the video and a manually
created mask using ImageJ software showing only the ventricle.
Following mask creation, all image and mask sets were organized into folders, with each
set containing two folders: one for the extracted original frame and the other for its
corresponding mask.
For the validation set, all annotated images from 4 of the 41 videos were used, thus
ensuring that the validation set is independent of the training set.
After this, the sets were randomly shuffled into validation and training data.

\subsubsection{Data Augmentation}

In semantic segmentation, data augmentation is a technique used to increase the size of
the training dataset by creating new training examples from the existing
ones~\cite{shorten2019}.
This is achieved by applying a range of transformations to the original training images,
resulting in new images that are still representative of the same underlying scene or
object, but with variations in appearance.
Data augmentation is commonly used in deep learning-based computer vision tasks,
including semantic segmentation, to prevent overfitting and improve the generalization
capability of the model~\cite{cossio2023}.

In the original ZACAF implementation no augmentation was used since all the videos were
recorded with a uniform placement under the camera~\cite{naderi2021}.
However, in this dataset the orientation of the fish was random.
This augmentation assures that the framework is less prone to the orientation of the fish
under the microscope and camera setup, which provides the user with more flexibility for
recording.
Additionally, data augmentation is one of the methods used for improving performance with
limited data and reducing overfitting.
Lastly, using data augmentation enables TTA, which is discussed in the next section.
Here, only horizontal and vertical flipping transformations have been used to imitate all
the possible fish placements during the recording process.
These transformations were used since they represented the variations naturally possible
by physically changing the displacement of the fish.

\subsubsection{Transfer Learning}

Transfer learning is a machine learning technique where a pre-trained model, typically
trained on a large dataset, is used as a starting point for a new task or
problem~\cite{pan2010}.
The pre-trained model has already learned a set of feature representations that are
applicable to a wide range of problems, and these learned features can be transferred
and fine-tuned to the new problem with a smaller dataset.
This allows for faster training and better performance than training a model from scratch
on the new dataset.
If TL is not utilized, the model will be initialized with random weights.
In this case, the original ZACAF model was trained on a dataset of zebrafish recorded
previously using a different microscope setup~\cite{naderi2021}.
The dataset included less mature fish and different mutant types.
However, the features learned during the original model's training for ventricle
segmentation from zebrafish are expected to improve the training for the new dataset.
Therefore, we employed the pre-trained weights from the original model for the new model.
In this study, we demonstrate the efficacy of TL by selectively introducing only half of
the new dataset to our model.
This deliberate approach serves to illustrate that even with a limited dataset, the model
can be enhanced to adapt to a new microscope setup, encompassing diverse mutant types and
video recording protocols.
The remaining half of the new dataset was utilized to train a separate model, allowing for
a comprehensive comparison with the model trained solely through TL.
The application of TL facilitates the seamless integration of ZACAF into diverse setups
and new transgenic lines by emerging researchers and imparts a valuable aspect of
generalizability to the model.
Through successive rounds of TL conducted by users, the model is expected to exhibit
significant improvements in both generalizability and accuracy.
Consequently, our paper establishes a platform for zebrafish cardiovascular researchers,
enabling them to automatically calculate cardiovascular parameters with a modest,
annotated dataset while iteratively refining the model with each use.

\subsubsection{Test Time Augmentation}

Test-time augmentation (TTA) is a technique used in computer vision, including semantic
segmentation, to improve the accuracy of models during inference.
In semantic segmentation, TTA involves applying various image transformations or
augmentations to the test images and feeding them through the model multiple times to
obtain an ensemble of predictions.
These predictions are then combined, typically by averaging or voting, to obtain a final
segmentation map.
TTA helps to account for variability in the test data and reduces the risk of overfitting
to the training data.
These transformations create multiple versions of the same image, which can be fed through
the model to obtain a diverse set of predictions.

\begin{figure}[htbp]
    \centering
    \includegraphics[width=\linewidth]{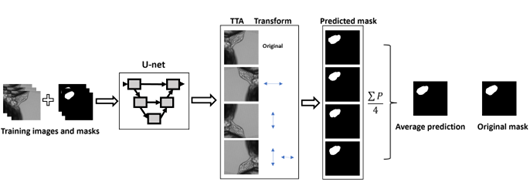}
    \caption{Implementation of the test time augmentation (TTA) techniques.
    The U-Net architecture is trained by an augmented dataset composed of images and their
    corresponding masks, and the TTA is applied to the test set output.
    The transformations used in TTA are horizontal and vertical flipping and their combination.
    These transformations, along with the original prediction, make 4 images which are then
    subjected to an element-wise average.
    On the far left of the figure the final prediction resulting from TTA can be compared with
    the original manually segmented mask.}
    \label{fig:fig2}
\end{figure}

Moshkov et al.\ incorporated TTA in the task of semantic segmentation of single-cell
analysis of microscopy images based on U-Net and Mask R-CNN deep learning models, which
showed improvement in the prediction accuracy~\cite{moshkov2020}.
A set of TTA techniques was then applied to the test image to generate multiple
predictions, which were subsequently combined to obtain a final prediction.
Specifically, horizontal flipping, vertical flipping, and a combination of both were
utilized to create three additional variations of the original test image.
For each of these variations, a prediction was obtained using the semantic segmentation model.
The four predictions were then combined by taking their element-wise average and
thresholding the result with a value of 0.2.
The thresholding is used to convert the soft masks resulting from element-wise averaging
into binary masks.
An example of TTA applied to a random image from the validation set can be seen in
Figure~\ref{fig:fig2}.

\subsubsection{Quantitative Comparison of Approaches}

Here, the same metrics used in ZACAF have been applied to evaluate the performance of
the deep learning model.
As the manually created masks are considered as the ground truth, the predicted shape
and manual mask are expected to be identical or similar.
Dice coefficient and Intersection over Union (IoU) are among the most used metrics in
semantic image segmentation.

\paragraph{a.~Dice coefficient.}
The Dice coefficient is a widely utilized metric to quantify the resemblance between two
objects, with a scale ranging from 0 to 1.
A value of 1 indicates a complete match or total overlap.
For a binary case, the coefficient is calculated as

\begin{equation}
    \text{Dice} = \frac{2\,|A \cap B|}{|A| + |B|}
    \label{eq:dice}
\end{equation}

where $A$ is the predicted image and $B$ is the ground truth (manually created mask).

\paragraph{b.~Intersection over Union.}
The Jaccard Index, also known as the Intersection over Union (IoU), measures the overlap
between the predicted segmentation and the ground truth by dividing their shared area by
their total area.
It ranges from 0 to 1, where 0 indicates no overlap and 1 indicates perfect overlap between
the two segmentations.
For the binary case, it can be calculated as:

\begin{equation}
    \text{IoU} = \frac{|A \cap B|}{|A \cup B|}
    \label{eq:iou}
\end{equation}

\section{Results}

\subsection{Assessment of the EF in \textit{Nrap}-Deficient Zebrafish}

\begin{figure}[htbp]
    \centering
    \includegraphics[width=0.50\linewidth]{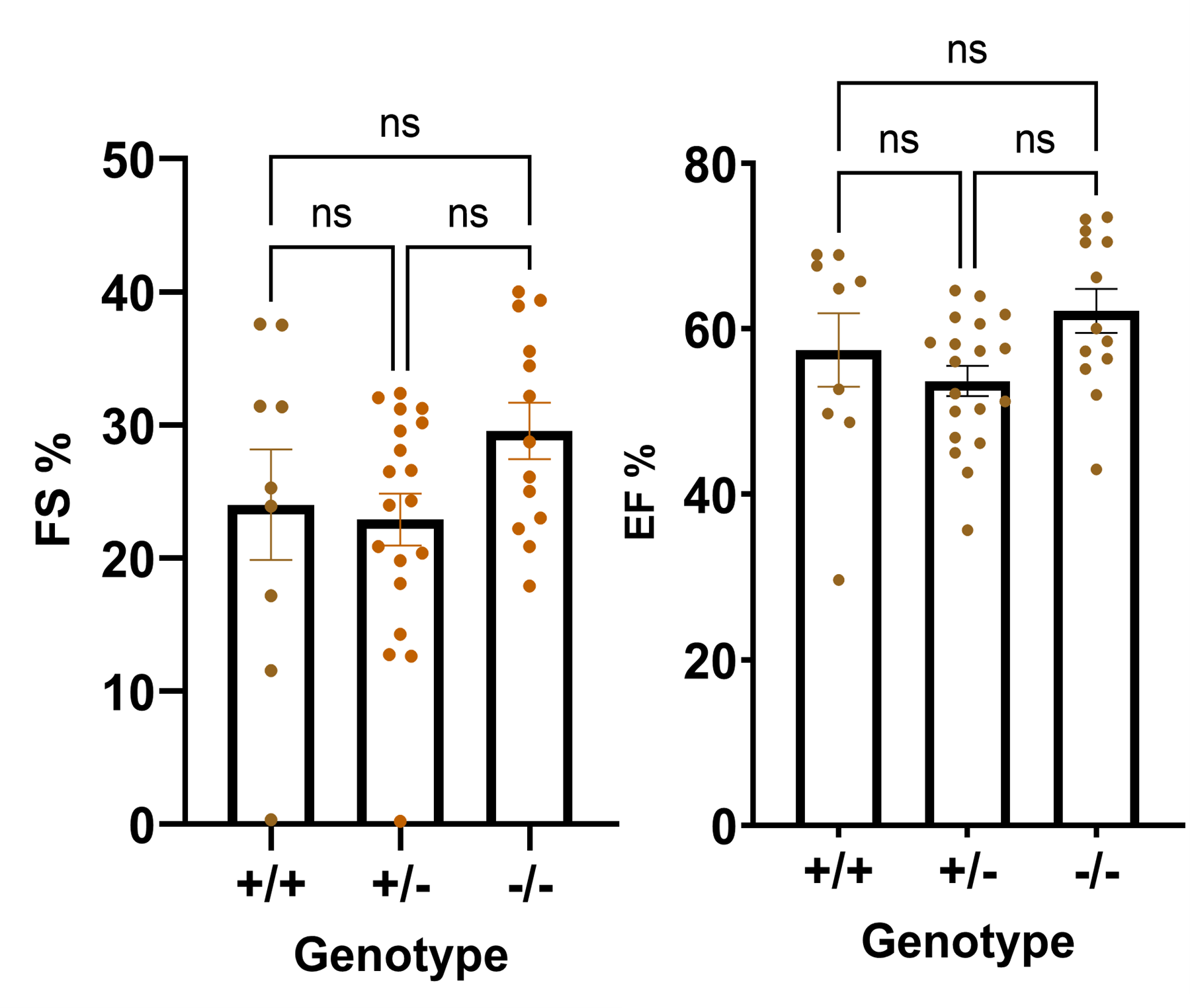}
    \caption{\textit{nrap} zebrafish model (2 dpf) EF and FS.
    Data represent mean $\pm$ s.e.m.\ with one-way ANOVA; ns, not significant.}
    \label{fig:fig3}
\end{figure}

Heart imaging in embryonic zebrafish (2.5 dpf, $n = 41$) was recorded and subsequent
genotyping revealed 19 \textit{nrap} heterozygotes (46\%), 9 wild type (22\%), and 13
mutants (31\%) in this group.
After the measurement of EF using the modified ZACAF, a one-way ANOVA test revealed no
significant differences between EF and FS in all three genotypes (Figure~\ref{fig:fig3}).
``Pear-shaped'' ventricles were observed during image processing in all genotypes
(18\% mutant, 36\% heterozygous, and 45\% wild type) (Figure~\ref{fig:fig1}),
yielding a frequency of 0.27.

\subsection{Assessment of the Performance of the Model with the Defined Metrics}

\begin{figure}[htbp]
    \centering
    \includegraphics[width=270px]{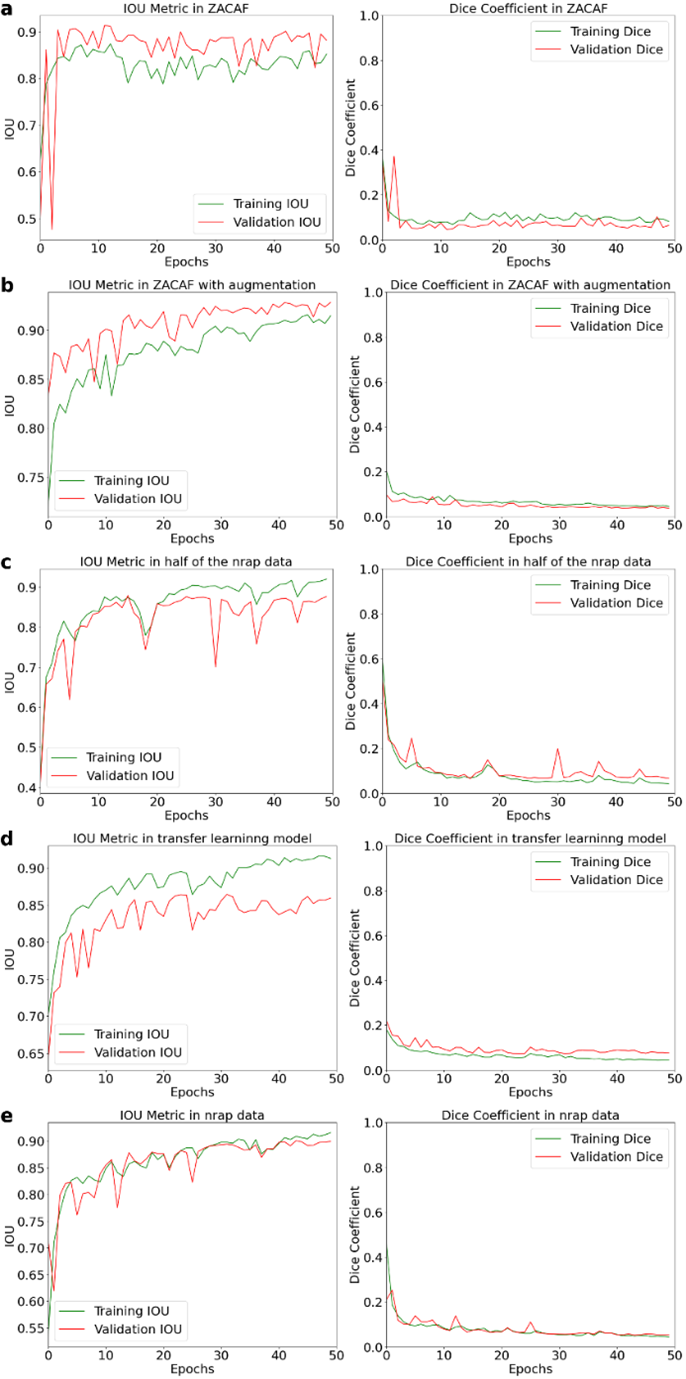}
    \caption{Comparison of the performance of the models in their Dice loss and IoU metric
    during training and validation.
    (a) ZACAF model trained on its original data.
    (b) ZACAF model trained on its original data with data augmentation.
    (c) Model trained on half of the \textit{nrap} data only.
    (d) Model trained on half of the \textit{nrap} data using TL by taking the ZACAF model
    pretrained weights.
    (e) Model trained only on the complete \textit{nrap} dataset.}
    \label{fig:fig4}
\end{figure}

The IoU metric serves as a performance assessment for the best-performing model, trained
utilizing a Dice loss function, an Adam optimizer, and a learning rate of 0.001 with decay
steps of 240 and a decay rate of 0.95.
The validation split encompassed 20\% or 80 sets.
The identical model underwent training on both ZACAF original data and the new \textit{nrap}
dataset, with and without the application of TL.
For TL, the training continued on the original ZACAF model using only half of the
\textit{nrap} data, demonstrating the efficacy of TL in the context of a limited dataset.
Figure~\ref{fig:fig4} depicts the training and validation Dice loss and IoU metric for five
different training regimens: ZACAF trained on its original data with and without data
augmentation, trained on the \textit{nrap} data, and ZACAF original fine-tuned on half of
the \textit{nrap} dataset.
To ensure model integrity, a model checkpoint was implemented as a callback to retain the
model with the highest validation IoU coefficient.

In Figure~\ref{fig:fig4}, the training and validation IoU rates for the original ZACAF were
88.1\% and 85.1\%, respectively, increasing to 92.4\% and 91.8\%, respectively, after
applying data augmentation to the training data.
Further, a minor overfitting is observed with the metrics being close.
This underscores the positive impact of data augmentation on the model.
Additionally, training the model on half of the \textit{nrap} data resulted in IoU rates
of 91.9\% and 87.6\% for training and validation, respectively.
Comparatively, employing TL on half of the data with the original ZACAF model yielded
rates of 91.6\% and 85.9\%, respectively.
Notably, the model trained solely with the complete \textit{nrap} dataset exhibited similar
performance.
The figure highlights that utilizing a small dataset (around 200 sets) with TL yielded
comparable results to a much larger \textit{nrap} dataset, emphasizing the potential of TL
as a framework for easy access to ZACAF while simultaneously enhancing its performance.

\subsection{Assessment of the Performance of the Framework for EF in the Test Set}

\begin{figure}[htbp]
    \centering
    \includegraphics[width=\linewidth]{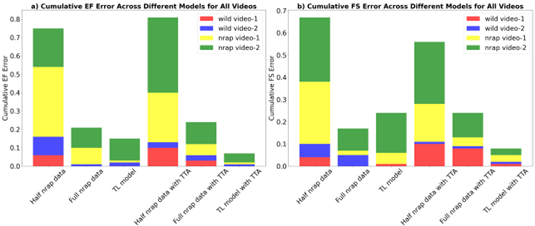}
    \caption{EF (a) and FS (b) were calculated from 4 test videos and the error with the
    manual calculation from all videos was accumulated.
    The error can be seen within different models; colors show the test video so the
    performance of each model can be seen for each video.
    From left to right on the $x$-axis: model trained on half of the \textit{nrap} data only;
    model trained on the complete \textit{nrap} dataset; model trained on half of the
    \textit{nrap} data using TL; and the same three models with TTA applied.
    The legend shows the contribution of each video to the cumulative error compared to
    manual calculation.}
    \label{fig:fig5}
\end{figure}

To evaluate the framework's effectiveness in calculating EF, we compared the results
obtained from manual assessment by an expert with those generated by the software.
As EF calculation requires finding the area in all frames of a video to determine the ED
and ES areas, the framework's performance was assessed using a series of frames from a
test video, rather than random images from a validation set.
To this end, we evaluated the framework's performance with EF calculation, using 2
wild-type zebrafish embryos and 2 \textit{nrap} mutant embryos as inputs for the test set,
without using them in training.
We first performed manual processing and estimation for each video to derive EF as the
ground truth.
Then, the model predicted ventricle masks for each frame of the input video, and the frames
with the maximum and minimum area of the segmented ventricle were identified as the ES and
ED frames, respectively.
It is worth mentioning that the outer edge of the ventricles was identified in the manual
segmentation process that created the training dataset; thus the model evaluation and results
were also based on that fact.
The framework subsequently computed EF and saved it, along with other indices, in a CSV file.
Among the five models discussed in the preceding section, three were selected for testing on
videos.
These models were trained on half of the \textit{nrap} data, trained on the full \textit{nrap}
data, and the TL model using half of the \textit{nrap} data on the original ZACAF weights.
Each of these models was independently employed to calculate three sets of EF and FS.
Additionally, TTA was applied to all three model predictions in a separate experiment,
resulting in a total of six sets of results.
In Figure~\ref{fig:fig5}, two distinct plots illustrate the cumulative error in four test
videos during the calculation of EF and FS.

\begin{figure}[htbp]
    \centering
    \includegraphics[width=0.85\linewidth]{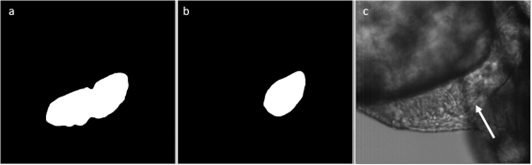}
    \caption{A frame from \textit{nrap} fish video recording where the ventricle and atrium
    are both visible.
    (a) Prediction without TTA.
    (b) Prediction after TTA.
    (c) Original video frame with an arrow showing the ventricle.}
    \label{fig:fig6}
\end{figure}

Firstly, the model trained with half of the \textit{nrap} data exhibited the highest error,
which was expected given the limited training data (200 sets) for a complex convolutional
neural network (CNN).
Conversely, the model trained on the complete \textit{nrap} dataset demonstrated significantly
lower error.
Secondly, the TL model, utilizing the original ZACAF model as pretrained weights and
continuing training with half of the \textit{nrap} dataset, showcased the effectiveness of TL
and the feasibility of utilizing such a small dataset.
This finding suggests that new users of ZACAF can leverage it by creating a small dataset.
Additionally, test videos from the original ZACAF dataset were tested using all models,
with only the TL model successfully segmenting the ventricle.
Notably, Figure~\ref{fig:fig5} highlights that TTA has substantially reduced the error.
TTA not only enhances edge accuracy but also proves beneficial when the model mistakenly
detects the atrium or other tissues.
Maintaining focus on the ventricle is challenging in the microscopic recording of zebrafish
heart.
TTA can address scenarios where both chambers are partially in focus, ensuring only the
ventricle is detected.
Figure~\ref{fig:fig6} provides an example of TTA's application in improving segmentation in
videos where both chambers are in focus.
Therefore, a notable advantage of employing TTA is that it facilitates the recording of
future datasets with greater ease, reducing the need for extensive cleaning and supervision.
The model incorporating TL and TTA demonstrated the best performance in the test set,
with average errors of 2\% and 1.7\% in EF and FS, respectively, in test videos.
Additionally, an intriguing observation from Figure~\ref{fig:fig5} indicates that mutant
\textit{nrap} fish exhibit higher error.

\section{Discussion}

We have developed a framework that can help researchers quantify the cardiac functions
and parameters of studied zebrafish with minimum manual engineering efforts.
In EF derivation, counting the pixels is more relevant and accurate than finding the
long axis, which can be complicated since the ventricle is not a perfect ellipse shape.
Further, most researchers use a freehand ruler in the ImageJ software, which could
introduce inaccuracy, especially with the small size of heart chambers.

The pear-shaped ventricles showed no significant correlation to the genotypes.
We speculate that this shape is observed due to the improper placement of the fish
under the microscope.
Considering that the ventricle is not a perfect ellipsoid, this shape could be the result
of the perspective of the camera from the ventricle, which can be imaged as a 2D shape
similar to a pear instead of being closer to an ellipse.

In Eq.~\ref{eq:vol_area}, $A$ is the 2D area calculated directly from the segmented ventricle,
and $D_L$ is the long axis.
This way of calculating the volume does not assume the shape of the ventricle to be a prolate
spheroid, unlike Eq.~\ref{eq:vol_prolate}.
This formula is useful specifically for mutant fish where the long and short axes might not
change significantly; however, the abnormal shape of the heart might contribute to an
abnormal EF measurement.
The results were calculated using both formulas.
The average difference between the EF measurements using the two formulas was 3.34\%,
which is negligible.
However, some videos showed significant differences of up to 19.5\%.

This unbiased analysis of the cardiac function revealed no significant differences in the
cardiac function in \textit{nrap} mutants compared to controls.
Recent human genetic studies and a study on the \textit{nrap} zebrafish model suggested
NRAP as a potential cause of cardiomyopathy.
The presence of the NRAP variant in unaffected family members or the presence of the mutant
phenotype also in the control zebrafish suggests a lack or extremely weak correlation of
NRAP as a cardiomyopathy-causing gene based on the previous findings~\cite{lu2011}.
Our previous studies identified improved skeletal muscle function through the reduction of
increased NRAP in nemaline myopathy.
This improved analysis of the cardiac function by the ZACAF models shows that NRAP reduction
is a potentially safe strategy in nemaline and related myopathies, as no cardiac defects were
observed in \textit{nrap} mutants during early development.

The use of TL in the ZACAF model has proved to be a successful technique for improving the
model's performance for ventricle segmentation in zebrafish.
By utilizing pre-trained weights from the original model, the new model was able to benefit
from the features learned during the previous training, resulting in faster training and better
performance than training the model from scratch on the new dataset.
Additionally, the use of callbacks, such as the model checkpoint, helped ensure that the
best-performing model was saved and used for further analysis, allowing the model to continue
improving its performance.

In this case, TTA was beneficial because it increased the variability of the test data and
helped to reduce the effect of any biases in the original dataset.
Since the ZACAF model was trained on a different dataset and microscope setup~\cite{naderi2021},
there could be some differences in the characteristics of the new dataset that were not
present in the original dataset.
By applying TTA, the model was able to generate additional test data that had different
characteristics, which helped to reduce the impact of any biases in the original dataset.
Another benefit of TTA is that it can help to increase the robustness of the model to
variations in the input data.
Since the model is exposed to a larger variety of test data during inference, it is less
likely to overfit to a particular type of input and more likely to generalize well to new data.
However, TTA also has some potential drawbacks.
One of the main drawbacks is that it can increase the computational cost of making predictions
since the model needs to process multiple versions of the test data.
The computational cost can be significant depending on the complexity of the model and the
number of test data versions generated.
Another potential drawback of TTA is that it can introduce some variability into the
predictions, which can make it difficult to interpret the results.
Since the final prediction is based on an average of multiple predictions, it may not be clear
which version of the test data was responsible for a particular prediction.
This can make it challenging to identify specific areas of the input data that the model is
struggling with.

\section{Conclusion}

Based on the findings presented in this paper, downregulation of Nebulin Related Anchoring
Protein (NRAP) in embryonic zebrafish does not significantly affect cardiac function,
specifically EF and FS, as measured by the Zebrafish Automatic Cardiovascular Assessment
Framework (ZACAF) based on a U-Net deep learning model.
While NRAP has been shown to play a role in myofibril assembly in cardiac and skeletal muscle,
and some human genetic studies have suggested an association of NRAP with cardiac function,
our results do not support the hypothesis that NRAP downregulation affects cardiac function.

The modified ZACAF deep learning model demonstrated effectiveness in implementing data from
new research teams with different recording setups and phenotypes and provided a fully automated
and accurate measurement of cardiovascular parameters, which could be useful in future research.
The effectiveness of data augmentation, TL, and TTA were evaluated.
Considerations for video recording and preprocessing, zebrafish orientation and handling, and
deep learning model architecture were discussed.

\section*{Acknowledgments}

The authors would like to acknowledge the financial support from the NIH STTR Phase I \#1R41HL166096 (H.C.), the NSF CAREER Award
\#1917105 (H.C.), R24 \#OD035402 (C.M.), and \#1K99HL161472-01 from NHLBI (D.C.).
Also, V.A.G.\ is supported by a grant from A Foundation Building Strength.

\section*{Declaration of Competing Interest}

The authors declare no competing interests.

\bibliographystyle{unsrt}
\bibliography{references}

@article{naderi2021,
  author  = {Naderi, Amir Mohammad and
             Bu, Haisong and
             Su, Jingcheng and
             Huang, Mao-Hsiang and
             Vo, Khuong and
             Trigo Torres, Ramses Seferino and
             Chiao, J-C and
             Lee, Juhyun and
             Lau, Michael P. H. and
             Xu, Xiaolei and
             Cao, Hung},
  title   = {Deep learning-based framework for cardiac function assessment
             in embryonic zebrafish from heart beating videos},
  journal = {Computers in Biology and Medicine},
  year    = {2021},
  volume  = {135},
  pages   = {104565},
  doi     = {10.1016/j.compbiomed.2021.104565}
}

@article{huang2023,
  author  = {Huang, Mao-Hsiang and
             Naderi, Amir Mohammad and
             Zhu, Ping and
             Xu, Xiaolei and
             Cao, Hung},
  title   = {Assessing Cardiac Functions of Zebrafish from Echocardiography
             Using Deep Learning},
  journal = {Information},
  year    = {2023},
  volume  = {14},
  number  = {6},
  pages   = {341},
  doi     = {10.3390/info14060341}
}

@article{wheeler2005,
  author  = {Wheeler, Ferrin C. and
             Fernandez, Liliana and
             Carlson, Kerri M. and
             Wolf, Matthew J. and
             Rockman, Howard A. and
             Marchuk, Douglas A.},
  title   = {{QTL} mapping in a mouse model of cardiomyopathy reveals an
             ancestral modifier allele affecting heart function and survival},
  journal = {Mammalian Genome},
  year    = {2005},
  volume  = {16},
  number  = {6},
  pages   = {414--423},
  doi     = {10.1007/s00335-005-2468-7}
}

@article{lu2008,
  author  = {Lu, Shajia and
             Borst, Diane E. and
             Horowits, Robert},
  title   = {Expression and alternative splicing of {N-RAP} during mouse
             skeletal muscle development},
  journal = {Cell Motility and the Cytoskeleton},
  year    = {2008},
  volume  = {65},
  number  = {12},
  pages   = {945--954},
  doi     = {10.1002/cm.20311}
}

@article{jirka2019a,
  author  = {Jirka, Caroline and
             Pak, Jasmine H. and
             Grosgogeat, Claire A. and
             Marchetti, Michael Mario and
             Gupta, Vandana A.},
  title   = {Dysregulation of {NRAP} degradation by {KLHL41} contributes
             to pathophysiology in nemaline myopathy},
  journal = {Human Molecular Genetics},
  year    = {2019},
  volume  = {28},
  number  = {15},
  pages   = {2549--2560},
  doi     = {10.1093/hmg/ddz078}
}

@article{lu2011,
  author  = {Lu, Shajia and
             Crawford, Garland L. and
             Dore, Justin and
             Anderson, Stasia A. and
             Despres, Daryl and
             Horowits, Robert},
  title   = {Cardiac-specific {NRAP} overexpression causes right ventricular
             dysfunction in mice},
  journal = {Experimental Cell Research},
  year    = {2011},
  volume  = {317},
  number  = {8},
  pages   = {1226--1237},
  doi     = {10.1016/j.yexcr.2011.01.020}
}

@article{truszkowska2017,
  author  = {Truszkowska, Gra{\.z}yna T. and
             Bili{\'n}ska, Zofia T. and
             Muchowicz, Angelika and
             Pollak, Agnieszka and
             Biernacka, Anna and
             Kozar-Kami{\'n}ska, Katarzyna and
             Stawi{\'n}ski, Piotr and
             Gasperowicz, Piotr and
             Kosi{\'n}ska, Joanna and
             Zieli{\'n}ski, Tomasz and
             P{\l}oski, Rafa{\l}},
  title   = {Homozygous truncating mutation in {NRAP} gene identified by
             whole exome sequencing in a patient with dilated cardiomyopathy},
  journal = {Scientific Reports},
  year    = {2017},
  volume  = {7},
  number  = {1},
  pages   = {3362},
  doi     = {10.1038/s41598-017-03189-8}
}

@article{koskenvuo2021,
  author  = {Koskenvuo, Juha W. and
             Saarinen, Inka and
             Ahonen, Saija and
             Tommiska, Johanna and
             Weckstr{\"o}m, Sini and
             Sepp{\"a}l{\"a}, Eija H. and
             Tuupanen, Sari and
             Kangas-Kontio, Tiia and
             Schleit, Jennifer and
             Heli{\"o}, Krista and
             Hathaway, Julie and
             Gummesson, Anders and
             Dahlberg, Pia and
             Ojala, Tiina H. and
             Veps{\"a}l{\"a}inen, Ville and
             Kyt{\"o}l{\"a}, Ville and
             Muona, Mikko and
             Sistonen, Johanna and
             Salmenper{\"a}, Pertteli and
             Gentile, Massimiliano and
             Paananen, Jussi and
             Myllykangas, Samuel and
             Alastalo, Tero-Pekka and
             Heli{\"o}, Tiina},
  title   = {Biallelic loss-of-function in {NRAP} is a cause of recessive
             dilated cardiomyopathy},
  journal = {{PLOS ONE}},
  year    = {2021},
  volume  = {16},
  number  = {2},
  pages   = {e0245681},
  doi     = {10.1371/journal.pone.0245681}
}

@article{zhang2023nrap,
  author  = {Zhang, Zhongman and
             Xu, Kangkang and
             Ji, Lianfu and
             Zhang, Han and
             Yin, Jie and
             Zhou, Ming and
             Wang, Chunli and
             Yang, Shiwei},
  title   = {A novel loss-of-function mutation in {NRAP} is associated with
             left ventricular non-compaction cardiomyopathy},
  journal = {Frontiers in Cardiovascular Medicine},
  year    = {2023},
  volume  = {10},
  pages   = {1097957},
  doi     = {10.3389/fcvm.2023.1097957}
}

@inproceedings{ronneberger2015,
  author    = {Ronneberger, Olaf and
               Fischer, Philipp and
               Brox, Thomas},
  title     = {{U-Net}: Convolutional Networks for Biomedical Image Segmentation},
  booktitle = {Medical Image Computing and Computer-Assisted Intervention --
               {MICCAI} 2015, 18th International Conference, Munich, Germany,
               October 5--9, 2015, Proceedings, Part {III}},
  series    = {Lecture Notes in Computer Science},
  volume    = {9351},
  pages     = {234--241},
  publisher = {Springer, Cham},
  year      = {2015},
  doi       = {10.1007/978-3-319-24574-4_28}
}

@inproceedings{long2015,
  author    = {Long, Jonathan and
               Shelhamer, Evan and
               Darrell, Trevor},
  title     = {Fully Convolutional Networks for Semantic Segmentation},
  booktitle = {Proceedings of the {IEEE} Conference on Computer Vision and
               Pattern Recognition ({CVPR})},
  pages     = {3431--3440},
  address   = {Boston, MA},
  year      = {2015},
  doi       = {10.1109/CVPR.2015.7298965}
}

@article{badrinarayanan2017,
  author  = {Badrinarayanan, Vijay and
             Kendall, Alex and
             Cipolla, Roberto},
  title   = {{SegNet}: A Deep Convolutional Encoder-Decoder Architecture
             for Image Segmentation},
  journal = {{IEEE} Transactions on Pattern Analysis and Machine Intelligence},
  year    = {2017},
  volume  = {39},
  number  = {12},
  pages   = {2481--2495},
  doi     = {10.1109/TPAMI.2016.2644615}
}

@article{wessells2004,
  author  = {Wessells, Robert J. and
             Bodmer, Rolf},
  title   = {Screening assays for heart function mutants in
             \textit{{Drosophila}}},
  journal = {BioTechniques},
  year    = {2004},
  volume  = {37},
  number  = {1},
  pages   = {58--66},
  doi     = {10.2144/04371ST01}
}

@article{nasrat2016,
  author  = {Nasrat, Sara and
             Marcato, Daniel and
             Hirth, Sofia and
             Reischl, Markus and
             Pylatiuk, Christian},
  title   = {Semi-automated detection of fractional shortening in zebrafish
             embryo heart videos},
  journal = {Current Directions in Biomedical Engineering},
  year    = {2016},
  volume  = {2},
  number  = {1},
  pages   = {233--236},
  doi     = {10.1515/cdbme-2016-0052}
}

@article{akerberg2019,
  author  = {Akerberg, Alexander A. and
             Burns, Caroline E. and
             Burns, C. Geoffrey and
             Nguyen, Christopher},
  title   = {Deep learning enables automated volumetric assessments of
             cardiac function in zebrafish},
  journal = {Disease Models \& Mechanisms},
  year    = {2019},
  volume  = {12},
  number  = {10},
  pages   = {dmm040188},
  doi     = {10.1242/dmm.040188}
}

@article{huang2000,
  author  = {Huang, W. Y. and
             Aramburu, J. and
             Douglas, P. S. and
             Izumo, S.},
  title   = {Transgenic expression of green fluorescence protein can cause
             dilated cardiomyopathy},
  journal = {Nature Medicine},
  year    = {2000},
  volume  = {6},
  number  = {5},
  pages   = {482--483},
  doi     = {10.1038/74914}
}

@article{zhang2021,
  author  = {Zhang, Bingyu and
             Pas, Kelsey E. and
             Ijaseun, Tomolemi and
             Cao, Hung and
             Fei, Peng and
             Lee, Juhyun},
  title   = {Automatic Segmentation and Cardiac Mechanics Analysis of
             Evolving Zebrafish Using Deep Learning},
  journal = {Frontiers in Cardiovascular Medicine},
  year    = {2021},
  volume  = {8},
  pages   = {675291},
  doi     = {10.3389/fcvm.2021.675291}
}

@article{suryanto2022,
  author  = {Suryanto, Michael Edbert and
             Saputra, Ferry and
             Kurnia, Kevin Adi and
             Vasquez, Ross D. and
             Roldan, Marri Jmelou M. and
             Chen, Kelvin H-C. and
             Huang, Jong-Chin and
             Hsiao, Chung-Der},
  title   = {Using {DeepLabCut} as a Real-Time and Markerless Tool for
             Cardiac Physiology Assessment in Zebrafish},
  journal = {Biology},
  year    = {2022},
  volume  = {11},
  number  = {8},
  pages   = {1243},
  doi     = {10.3390/biology11081243}
}

@book{westerfield1995,
  author    = {Westerfield, Monte},
  title     = {The Zebrafish Book: A Guide for the Laboratory Use of Zebrafish
               (\textit{{Danio rerio}})},
  edition   = {3rd},
  publisher = {University of Oregon Press},
  address   = {Eugene, OR},
  year      = {1995}
}

@article{casey2023,
  author  = {Casey, Jennifer G. and
             Kim, Euri S. and
             Joseph, Remi and
             Li, Frank and
             Granzier, Henk and
             Gupta, Vandana A.},
  title   = {{NRAP} reduction rescues sarcomere defects in nebulin-related
             nemaline myopathy},
  journal = {Human Molecular Genetics},
  year    = {2023},
  volume  = {32},
  number  = {10},
  pages   = {1711--1721},
  doi     = {10.1093/hmg/ddad011}
}

@article{kimmel1995,
  author  = {Kimmel, Charles B. and
             Ballard, William W. and
             Kimmel, Seth R. and
             Ullmann, Bonnie and
             Schilling, Thomas F.},
  title   = {Stages of embryonic development of the zebrafish},
  journal = {Developmental Dynamics},
  year    = {1995},
  volume  = {203},
  number  = {3},
  pages   = {253--310},
  doi     = {10.1002/aja.1002030302}
}

@article{yalcin2017,
  author  = {Yalcin, Huseyin C. and
             Amindari, Arbia and
             Butcher, Jonathan T. and
             Althani, Asma and
             Yacoub, Magdi},
  title   = {Heart function and hemodynamic analysis for zebrafish embryos},
  journal = {Developmental Dynamics},
  year    = {2017},
  volume  = {246},
  number  = {11},
  pages   = {868--880},
  doi     = {10.1002/dvdy.24497}
}

@article{wisneski1981,
  author  = {Wisneski, J. A. and
             Pfeil, C. N. and
             Wyse, D. G. and
             Mitchell, R. and
             Rahimtoola, S. H. and
             Gertz, E. W.},
  title   = {Left ventricular ejection fraction calculated from volumes
             and areas: underestimation by area method},
  journal = {Circulation},
  year    = {1981},
  volume  = {63},
  number  = {1},
  pages   = {149--151},
  doi     = {10.1161/01.CIR.63.1.149}
}

@article{shorten2019,
  author  = {Shorten, Connor and
             Khoshgoftaar, Taghi M.},
  title   = {A survey on Image Data Augmentation for Deep Learning},
  journal = {Journal of Big Data},
  year    = {2019},
  volume  = {6},
  number  = {1},
  pages   = {60},
  doi     = {10.1186/s40537-019-0197-0}
}

@article{cossio2023,
  author  = {Cossio, Manuel},
  title   = {Augmenting Medical Imaging: A Comprehensive Catalogue of 65
             Techniques for Enhanced Data Analysis},
  journal = {arXiv preprint},
  year    = {2023},
  note    = {arXiv:2303.01178}
}

@article{pan2010,
  author  = {Pan, Sinno Jialin and
             Yang, Qiang},
  title   = {A Survey on Transfer Learning},
  journal = {{IEEE} Transactions on Knowledge and Data Engineering},
  year    = {2010},
  volume  = {22},
  number  = {10},
  pages   = {1345--1359},
  doi     = {10.1109/TKDE.2009.191}
}

@article{moshkov2020,
  author  = {Moshkov, Nikita and
             Mathe, Botond and
             Kertesz-Farkas, Attila and
             Hollandi, Reka and
             Horvath, Peter},
  title   = {Test-time augmentation for deep learning-based cell segmentation
             on microscopy images},
  journal = {Scientific Reports},
  year    = {2020},
  volume  = {10},
  number  = {1},
  pages   = {5068},
  doi     = {10.1038/s41598-020-61808-3}
}

\end{document}